\documentclass[12pt]{iopart}

\usepackage{graphicx}
\usepackage{dcolumn}
\usepackage{bm}

\makeatletter
\@ifundefined{showcaptionsetup}{}{%
\PassOptionsToPackage{caption=false}{subfig}}
\usepackage{subfig}
\makeatother

\begin{document}

\title{High-Fidelity State Detection and Tomography of a Single Ion Zeeman Qubit}
\author{A Keselman, Y Glickman, N Akerman, S Kotler and R Ozeri}
\address{Physics of Complex Systems, The Weizmann Institute of Science, Rehovot 76100, Israel}

\begin{abstract}
We demonstrate high-fidelity Zeeman qubit state detection in a
single trapped $^{88}Sr^{+}$ ion. Qubit readout is performed by
shelving one of the qubit states to a metastable level using a
narrow linewidth diode laser at $674\ nm$ followed by
state-selective fluorescence detection. The average fidelity reached
for the readout of the qubit state is 0.9989(1). We then measure the
fidelity of state tomography, averaged over all possible
single-qubit states, which is 0.9979(2). We also fully characterize the
detection process using quantum process tomography. This readout
fidelity is compatible with recent estimates of the detection error-threshold required
for fault-tolerant computation, whereas high-fidelity state tomography opens the way for high-precision
quantum process tomography.
\end{abstract}

\maketitle

\section{Introduction}

One of the basic requirements for implementing a physical qubit is
the ability to faithfully measure its state
\cite{divincenzo2001physical}. Furthermore, qubit state detection
must be performed with high fidelity to reach fault tolerant
quantum computation. The exact detection error threshold required depends on many factors;
however under quite general assumptions error values on the order of
$10^{-2}-10^{-4}$ were estimated \cite{knill2005quantum}. Another
important use of high-fidelity state detection is precision
process tomography for studying different quantum processes.

Trapped ion qubits are a promising candidate system for physically
realizing a quantum computer and serve as a convenient test-ground
for studying fundamental quantum dynamics. Ion species used for this purpose
typically have a single electron in their valence shell and
the two qubit states are encoded in two energy levels of the valence
electron. State detection methods used so far rely on
state-selective fluorescence. Here photons are scattered from a
laser that is resonant with a transition from one of the qubit
states to a short-lived excited state, whereas transitions from the
other qubit state are largely off-resonance. State inference is then
based on the detected photon statistics \cite{bible}. To this end,
qubit choices with a large energy separation are advantageous.

For optical qubits, in which the two qubit states are separated by
an optical transition, state detection fidelity as high as 0.9999
was demonstrated using state-selective florescence and accounting
for photon arrival times \cite{myerson2008high,Burrell2010multi}. Another detection
scheme, using repetitive quantum nondemolition measurements on an
ancila ion-qubit, was shown to give a fidelity of 0.9994
\cite{qnd_detection-hume-wineland}. Optical qubits, however, have the
disadvantage of an excited state lifetime on the order of one
second, depending on the exact ion species used. Furthermore, the
linewidth of even ultra-stable optical local oscillators, i.e.,
frequency stabilized lasers, would limit the dephasing time of an
optical qubit.

Qubits that are encoded into a pair of levels in the electronic
ground state, in which the two qubit levels are split, either by the
Zeeman effect or the hyperfine interaction, by radio-frequency
transitions, have practically an infinite lifetime, as well as a very
long coherence time \cite{langer2005long}. Hyperfine qubit levels
are typically separated by frequencies in the GHz range. State
detection for this type of qubit can still be implemented by direct
state-selective fluorescence, since the typical linewidth of
electric-dipole transitions is two orders of magnitude smaller.
Here, off-resonance scattering normally limits state detection
fidelity to below 0.995
\cite{cd+hyperfine_ccd,langerthesis,yb+hyperfine_detection_monroe}.
The use of ancila qubits can, in principle, increase the detection
fidelity of a hyperfine ion-qubit \cite{SchaetzPRL}. In ion species
that have low-lying meta-stable levels, one of the qubit states can
be shelved to a meta-stable level prior to detection. Here detection
fidelity is similar to that of an optical qubit with an additional
error introduced by the state shelving process. Using state
selective optical pumping for shelving, a hyperfine ion-qubit
measurement fidelity as high as 0.9977 was demonstrated.
\cite{myerson2008high}.

Ion-qubits that are encoded into a pair of Zeeman split levels pose
the hardest state-measurement challenge. This is because the
frequency separation between the qubit levels is typically in the
MHz range and is comparable to the spectral linewidth of
electric-dipole transitions used for state-selective fluorescence.
Since direct state-selective fluorescence is impossible, Zeeman qubits
can be readout only by state mapping onto a Hyperfine ancila qubit
\cite{qnd_detection-hume-wineland} or by shelving one of the qubit
states onto a meta-stable level. However, the fidelity of state shelving in a
Zeeman qubit using optical pumping is limited due to the strong
coupling of the shelving light to the other qubit state
\cite{Oxford2004}. Thus, the only way to state-selectively shelve
a Zeeman qubit with high fidelity is by using a narrow-linewidth laser.
Although there have been several reports on detecting a
Zeeman qubit with shelving via a narrow-linewidth laser, with
typical fidelities below 0.996
\cite{wunderlich2007rap,schmidt-kaler2009zeemanCa}, to our
knowledge, there hasn't been a systematic study of the measurement
error and limitations in this kind of qubit.

Here we demonstrate the readout of a single-ion Zeeman qubit with
a fidelity of 0.9989(1). Spin state detection is performed via
electron shelving with a narrow-linewidth diode laser, followed by
state-selective fluorescence. We analyze the different fundamental sources as
well as technical sources of measurement error in detail. Further,
we use the high-fidelity state detection of our qubit to demonstrate
high-fidelity state tomography \cite{nielsenchuang}. Here, we
measure the fidelity of quantum state tomography averaged over all
possible single qubit states and also fully characterize
the detection process using quantum process tomography.

\section{Experimental Setup}

\begin{figure}[htb]
\begin{center}
  \includegraphics[trim = 25pt 30pt 35pt 20pt, clip, width=0.6\textwidth]{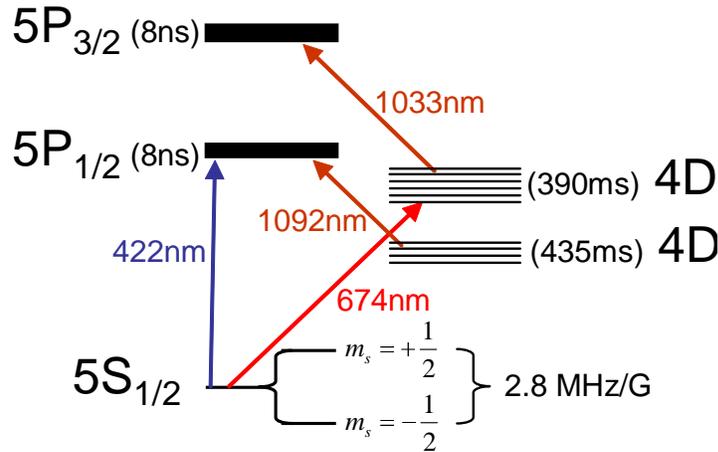}
  \caption{Energy levels scheme of the single valence electron of the $^{88}Sr^{+}$ ion.
  The $|\uparrow\rangle$ and $|\downarrow\rangle$ qubit states are encoded in the two,
  Zeeman-split, spin $1/2$ states of the $S_{1/2}$ ground level.
  Energy level lifetimes are written next to their spectroscopic notation.
  Laser light at 422 nm performs laser-cooling and state-selective fluorescence.
  Lasers at 1092 nm and 1033 nm pump out population from the meta-stable $D_{3/2}$ and $D_{5/2}$ levels respectively.
  A 674 nm narrow linewidth diode-laser shelves the electron from the qubit levels to levels in the $D_{5/2}$ manifold.}
  \label{fig:EnergyLevels}
\end{center}
\end{figure}

We trap a single $^{88}Sr^{+}$ ion in a linear RF Paul trap. The
trapping potential is well approximated as harmonic with a secular
axial frequency of $\omega_{ax}=(2\pi)1.09\ MHz$, and two nearly
degenerate radial frequencies of $\omega_{r}=(2\pi)2.5\ MHz$. A
scheme of the relevant energy levels in $^{88}Sr^+$ is shown in Fig.
\ref{fig:EnergyLevels}. The $|\uparrow\rangle$ and
$|\downarrow\rangle$ qubit states are encoded in the $5S_{1/2,
+1/2}$ and $5S_{1/2,-1/2}$ spin states respectively. A magnetic
field of $0.477$ mT splits the two qubit levels by a frequency,
$\omega_0 = (2\pi)13.366$ MHz.

An oscillating magnetic field, perpendicular to the quantization
axis, is generated by a current, oscillating at $\omega_0/(2\pi)$,
through an electrode positioned 2 mm from the ion, resulting in
coherent coupling between the two qubit levels. Zeeman qubit
coherent rotations are therefore generated by oscillating current
pulses.

In the first part of state detection the
electron is shelved from the $|\uparrow\rangle$ state to one of the $D_{5/2}$,
Zeeman-split, levels using a narrow linewidth, 674 nm, external
cavity diode laser. The diode laser is stabilized to a high finesse
($10^{5}$), ultra-low expansion glass, reference cavity. Residual
frequency noise of the laser has three dominant spectral features.
First, slow thermal drifts of the cavity result in drifts of the laser
frequency with a typical magnitude of 10 Hz/S. The atomic resonance
frequency is scanned every two minutes to correct for this drift.
Second, intermediate frequency noise results in a laser line half-width of
70 Hz over several seconds, estimated in a Ramsey spectroscopy
experiment \cite{Sengtock1994}. The third spectral feature results from fast
frequency noise that is unsuppressed, or even slightly enhanced, by
our frequency servo system. This spectral feature is often referred
to as the servo bump. Our servo bumps are centered around $700\ kHz$
on both sides of the carrier, having a width of $\sim300\ kHz$ , and
contain an estimated $<5\%$ of the optical power. A detailed
discussion of our narrow linewidth diode laser system is given in
\cite{Yinnon's paper}.
In this experiment, the direction of 674 nm laser light propagation relative to the
Zeeman splitting magnetic field direction, limits the allowed transitions between
$S_{1/2}$ and $D_{5/2}$ manifolds to $\Delta m = \pm 1$ transitions.

Electron shelving is followed by state-selective fluorescence
detection on the $S_{1/2}\rightarrow P_{1/2}$ transition at $422\
nm$. Photons that are scattered in a direction perpendicular to the
422 nm laser beam and the magnetic field are collected by an imaging
system (N.A. = 0.31) and are detected with a Photo Multiplier Tube
(PMT). State inference relies on the detected photon statistics.
Qualitatively, a small number of detected photons implies that the ion
qubit started in the $|\uparrow\rangle$ state and was shelved to the
non-fluorescing $D_{5/2}$ level, whereas a large number of detected photons
implies that the ion qubit started in the $|\downarrow\rangle$ state and
therefore was not shelved and remained at the fluorescing $S_{1/2}$ level.

\section{State Discrimination}

Following electron shelving, the number of photons $n$, detected by the PMT during a given detection time, $t_{det}$, is a random variable. This random variable is denoted by $n_{b}$ if the ion is in the fluorescing (bright) state $S_{1/2}$, and $n_{d}$ if the ion is in the non-fluorescing (dark) state $D_{5/2}$. Photon detection events, which occur when the ion is in the dark state are primarily due to scattering of the laser beam from trap surfaces. The fidelity of state discrimination is compromised by the overlap of the probability distribution functions (PDFs) of these two random variables. State inference can be performed by introducing a threshold value for the number of photons detected, $n_{th}$. If the number of photons detected is greater (smaller) than this threshold, $n > (\leq) n_{th}$, then we can infer that the ion is in the bright (dark) state. Given the probability distribution functions for $n_{b}$ and $n_{d}$, the errors in detecting the bright and dark states are $\epsilon_{b}=p_{b}\left(n\leq n_{th}\right)$ and $\epsilon_{d}=p_{d}\left(n>n_{th}\right)$ respectively. We want to find the parameters $t_{det}$ and $n_{th}$ that minimize the mean error
\begin{equation}
\epsilon=\frac{\epsilon_{b}+\epsilon_{d}}{2}=\frac{p_{b}\left(n\leq n_{th}\right)+p_{d}\left(n>n_{th}\right)}{2}. \label{eq:MeanError}
\end{equation}

\begin{figure}[htb]
\begin{center}
    \includegraphics[width=0.5\textwidth]{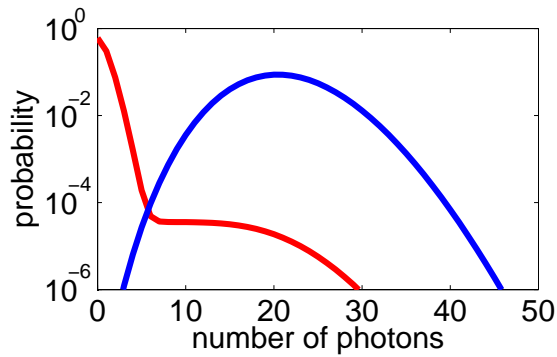}
    \caption{Theoretical probability density functions for the number of
    detected photons in the bright (blue line) and dark (red line)
    states, given in Eqs. \ref{eq:PDFBright} and \ref{eq:PDFDark}, using
    our measured photon detection rates $R_{b}=73.5kHz$ and
    $R_{d}=1.75kHz$, a detection time of $t_{det}=285\ \mu s$ and the known
    $D_{5/2}$ level-lifetime, $\tau_{D_{5/2}} = 390\ ms$.}
    \label{fig:HistPerfect}
\end{center}
\end{figure}

The detection fidelity is then given by $F=1-\epsilon$. If the
lifetime of the $D_{5/2}$ level would have been infinite, the random
variables $n_{b}$ and $n_{d}$ would follow two Poisson
distributions. Given photon detection rates $R_{b}$ and $R_{d}$ in
the bright and dark states respectively and a detection time,
$t_{det}$, the means of these distributions would be given by
$\bar{n}_{b,d}=R_{b,d}t_{det}$. Here the longer the detection time,
the smaller is the overlap between the two PDFs and therefore also
the detection error. However, the finite lifetime, $\tau_{D_{5/2}}$,
of the $D_{5/2}$ level introduces a correction to the PDF for
$n_{d}$, since there is a finite probability for the ion to decay
during the detection. Upon decay, the photon detection rate becomes
$R_{b}$. At a detection time much shorter than the $D_{5/2}$
lifetime, $t_{det} \ll \tau_{D_{5/2}}$, the PDFs for $n_{b}$ and
$n_{d}$ are given by \cite{mythesis} \numparts
\begin{eqnarray}
\label{eq:PDFBright}
\fl \qquad p_{b}\left(n\right)= Poiss\left(n,\:\bar{n}_{b}\right),\\
\label{eq:PDFDark}
\fl \qquad p_{d}\left(n\right)= \left(1-\frac{t_{det}}{\tau}\right)Poiss\left(n,\:\bar{n}_{d}\right)+ \frac{t_{det}}{\tau}\frac{\Gamma\left(\bar{n}_{b},\: n+1\right)-\Gamma\left(\bar{n}_{d},\: n+1\right)}{\bar{n}_{b}-\bar{n}_{d}},
\end{eqnarray}
\endnumparts
respectively. Here $Poiss\left(n,\:\bar{n}\right)$ denotes the
probability of detecting $n$ photons for a Poisson distribution with a
mean $\bar{n}$, and
$\Gamma\left(x,a\right)=\frac{1}{\Gamma\left(a\right)}\int_{0}^{x}e^{-t}t^{a-1}dt$
is the incomplete gamma function. The two PDFs are denoted by the blue
and red curves in Fig. \ref{fig:HistPerfect}, for our measured
photon detection rates $R_{b}=73.5kHz$ and $R_{d}=1.75kHz$, a
detection time of $t_{det}=285\ \mu s$ and the known $D_{5/2}$ level
lifetime, $\tau_{D_{5/2}} = 390\: ms$ \cite{sinclair2005lifetime}.

\begin{figure}[htb]
\begin{center}
\subfloat[\label{fig:TheorError2D}]
{\includegraphics[width=0.5\textwidth]{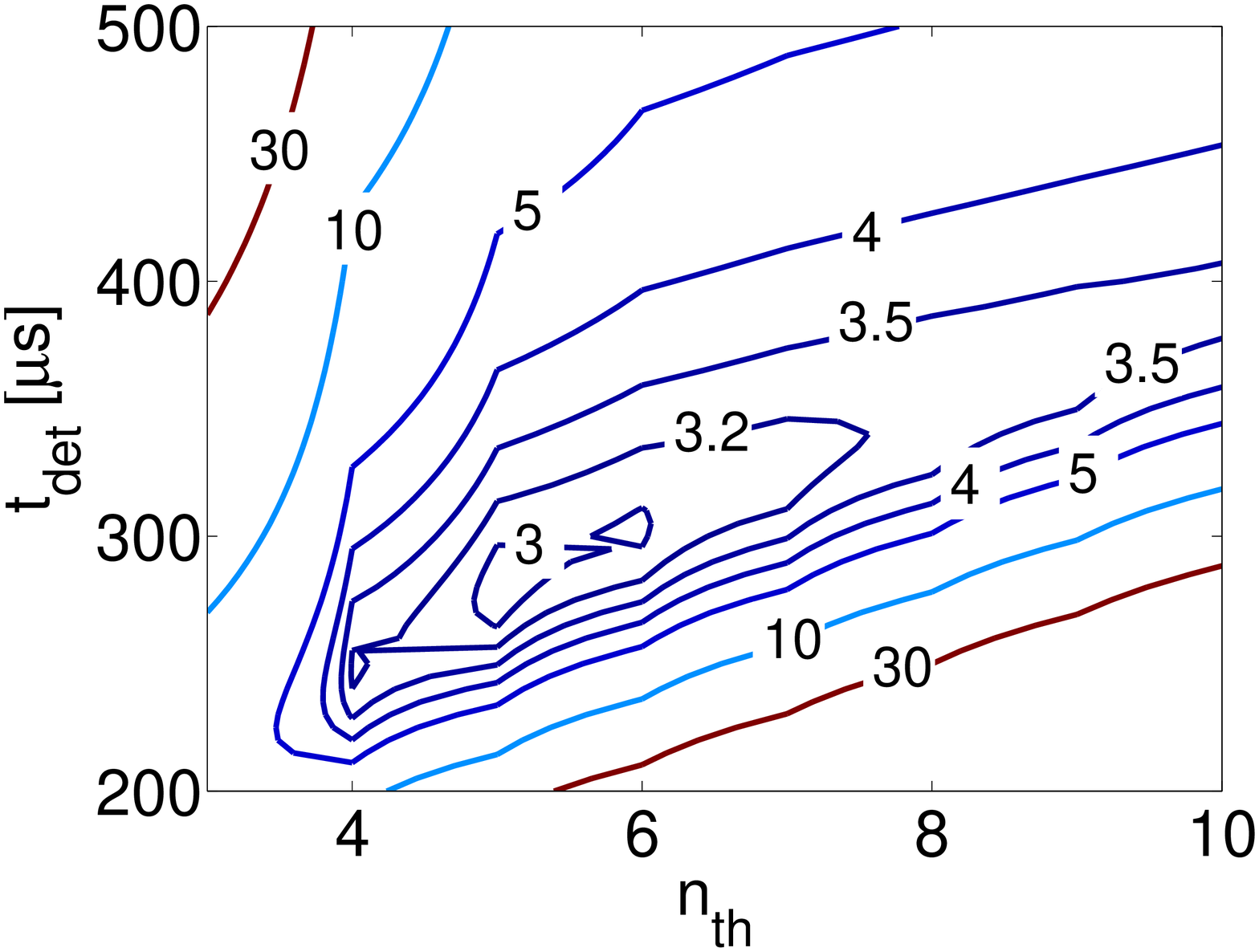}}
\subfloat[\label{fig:ExpError2D}]
{\includegraphics[width=0.5\textwidth]{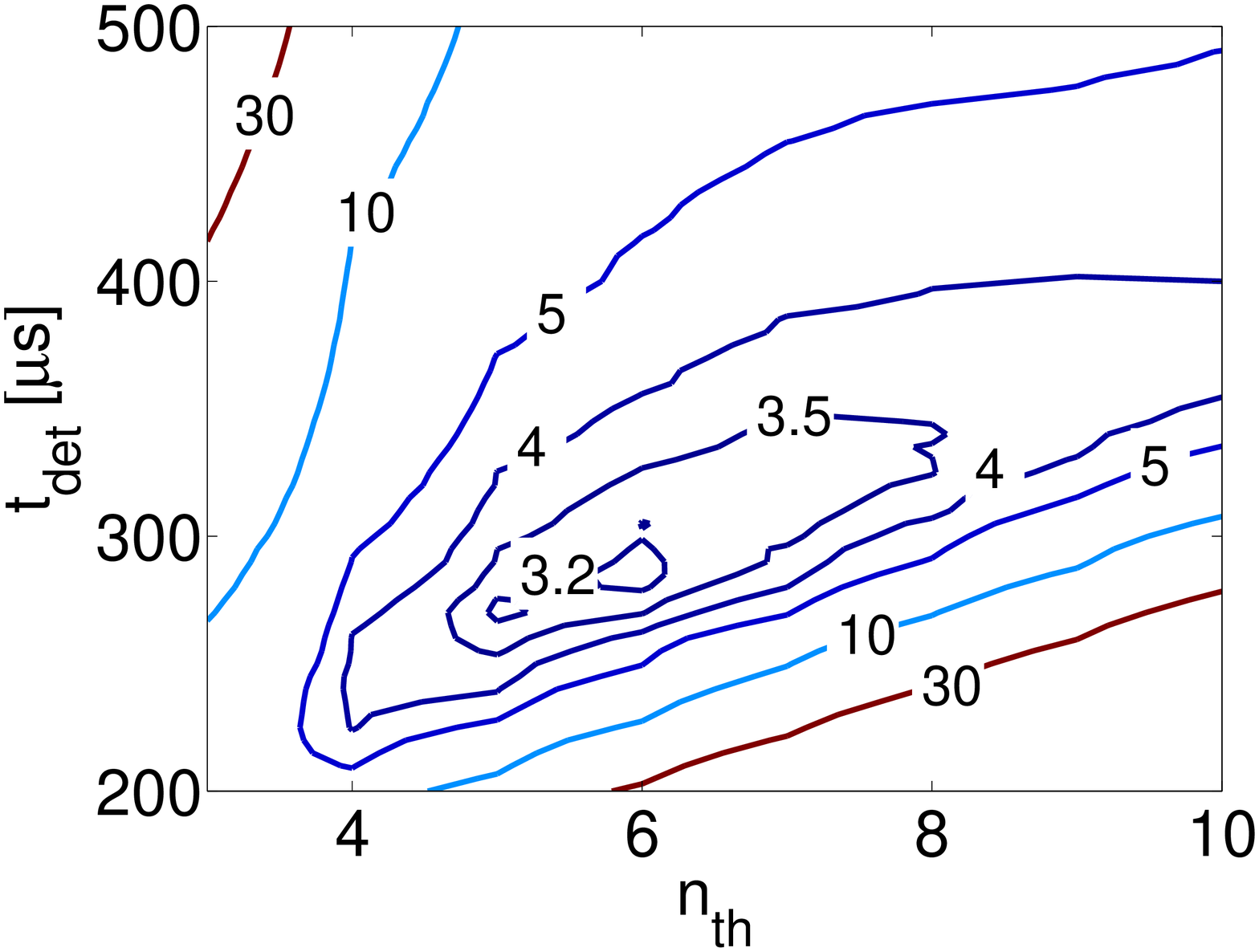}} 
\caption{State discrimination error as a function of the detection time $t_{det}$ and
the threshold on the number of photons $n_{th}$. (a) Theoretical
estimation obtained using Eqs. \ref{eq:PDFBright}, \ref{eq:PDFDark}.
The measured photon detection rates of $R_{b}=73.5kHz$ $R_{d}=1.75kHz$
are assumed, as well as the known lifetime of the $D_{5/2}$ level,
$\tau_{D_{5/2}}=390\ ms$ \cite{sinclair2005lifetime}. A minimal
error of $\epsilon=2.9\cdot10^{-4}$ is calculated at a detection
time of $t_{det}=280\mu s$ and a $n_{th}=5$ threshold value for the number of
detected photons. (b) Experimental results. Here the mean
initialization and shelving error of $8\cdot10^{-4}$ was subtracted.
A minimal error of $\epsilon=3(1)\cdot10^{-4}$ is measured at a
detection time of $t_{det}=285\mu s$ and a threshold value for the
number of photons of $n_{th}=6$. The theoretical model and our data
are seen to be in relatively good agreement.}
\end{center}
\end{figure}

As the detection time is increased, the overlap between the two
functions initially decreases owing to the larger spacing between the
two Poisson peaks, but it eventually increases owing to the growing tail
of the dark distribution. Hence, an optimal detection time and
a threshold number of photons exist such that the error in state
discrimination is minimal. Figure \ref{fig:TheorError2D} shows a
contour plot for $\epsilon$, as a function of the detection time
$t_{det}$ and the threshold on the number of photons $n_{th}$, for
the same $R_{b}$, $R_{d}$ and $\tau_{D_{5/2}}$ values used in Fig.
\ref{fig:HistPerfect}. A minimal error of
$\epsilon\simeq2.9\cdot10^{-4}$ is calculated at a detection time of
$t_{det}=280\mu s$ and a $n_{th}=5$ threshold value for the number of detected photons.

The actual distributions measured in the experiment are also
affected by the state preparation error, and an error resulting from an
imperfect shelving of the ion to the metastable state. State
preparation errors, $\epsilon_{\downarrow,init}$ and
$\epsilon_{\uparrow,init}$ for the $|\downarrow\rangle$ and $|\uparrow\rangle$
states, respectively, are given by the fraction of experiments in which the
ion was initialized in the wrong state. The shelving error for the
$|\uparrow\rangle$ state, $\epsilon_{\uparrow,shelving}$, is the
probability that the ion remained at the $S_{1/2}$ level after
shelving was performed. For the $|\downarrow\rangle$ state,
$\epsilon_{\downarrow,shelving}$ is the probability that the ion was
shelved to the $D_{5/2}$ level due to off-resonant light. Neglecting
terms that are second order in the different errors, the resulting
PDFs for the $|\downarrow\rangle$ and $|\uparrow\rangle$ states are
given by \cite{mythesis}, \numparts
\begin{eqnarray}
\fl \qquad \tilde{p}_{\downarrow}\left(n\right) \simeq \left(1-\epsilon_{\downarrow,tot}\right)Poiss\left(n,\bar{n}_{b}\right) +\epsilon_{\downarrow,tot}Poiss\left(n,\bar{n}_{d}\right)
\label{eq:PDFWithErrorBrightApp} \\
\fl \qquad \tilde{p}_{\uparrow}\left(n\right) \simeq \left(1-\epsilon_{\uparrow,tot}- \frac{t_{det}}{\tau}\right)Poiss\left(n,\:\bar{n}_{d}\right)+ \nonumber \\
\frac{t_{det}}{\tau}\frac{\Gamma\left(\bar{n}_{b}\: n+1\right)-\Gamma\left(\bar{n}_{d},\: n+1\right)}{\bar{n}_{b}-\bar{n}_{d}}+\epsilon_{\uparrow,tot}Poiss\left(n,\bar{n}_{b}\right).
\label{eq:PDFWithErrorDarkApp}
\end{eqnarray}
\endnumparts

Note that only the sum of the initialization error and the shelving error appears
$\epsilon_{\downarrow/\uparrow,tot}=\epsilon_{\downarrow/\uparrow,init}+\epsilon_{\downarrow/\uparrow,shelving}$.
This prevents us from distinguishing the state preparation error from the shelving error. The total mean detection error,
\begin{equation}
\tilde{\epsilon}=\frac{\tilde{\epsilon}_{\downarrow}+\tilde{\epsilon}_{\uparrow}}{2}=\frac{\tilde{p}_{\downarrow}\left(n\leq n_{c}\right)+\tilde{p}_{\uparrow}\left(n>n_{c}\right)}{2},
\end{equation}
can be related to the error $\epsilon$ resulting from the finite lifetime of the metastable level alone, obtained before
\begin{equation}
\tilde{\epsilon}=\epsilon+\frac{\epsilon_{\downarrow,tot}+\epsilon_{\uparrow,tot}}{2}.
\label{eq:TotalError}
\end{equation}
In particular, a minimum of $\tilde{\epsilon}$ is obtained for the same values of detection time $t_{det}$ and
the threshold on the number of photons $n_{th}$, as the minimum of $\epsilon$.

\section{Experimental Sequence}

In the experiment, two sets of data are taken. In each set the ion is prepared in one of the two Zeeman qubit
states, $|\uparrow\rangle$ or $|\downarrow\rangle$, and then state detection is performed. Each set of data
contains $3\cdot10^{5}$ repetitions of the experiment, yielding statistical uncertainty of the estimated
measurement error which is below $1\cdot10^{-4}$.

In both sequences the ion is first Doppler cooled on the
$S_{1/2}\rightarrow P_{1/2}$ transition for $300\ \mu s$, resulting
in a mean axial harmonic oscillator number of $\bar{n}\simeq25$.
Second, sideband cooling is performed on the $S_{1/2,+1/2}
\rightarrow D_{5/2,+3/2}$ narrow transition. To this end, the red
sideband of the $S_{1/2,+1/2}\rightarrow D_{5/2,+3/2}$ transition
is continuously excited for $5\ ms$  with the $674\ nm$ laser, while
the $1033\ nm$ laser is left on to repump the population from the
$D_{5/2}$ metastable level, and a $\sigma^{+}$ polarized $422\ nm$
light is left on to repump the population from the $S_{1/2,-1/2}$ state
via the $P_{1/2}$ manifold. Following sideband cooling the mean
axial harmonic oscillator number is $\bar{n}= 0.3(2)$.

Following cooling, the spin state is initialized. At this
stage, it is highly probable that the qubit is already in the $|\uparrow\rangle$ state.
To increase this probability further, we leave the
$\sigma^{+}$ polarized, $422\ nm$, optical pumping beam for an
additional $50\ \mu s$. To initialize the qubit in the
$|\downarrow\rangle$ state, we use an $8\ \mu s$ coherent qubit rotation
to bring the electron from $|\uparrow\rangle$ to $|\downarrow\rangle$.
Optical pumping using
the $674\ nm$ laser follows, to further increase the initialization efficiency.
Ten consecutive $\pi$-pulses on the $S_{1/2,\mp1/2}\rightarrow D_{5/2,\pm1/2}$ transition, each followed
by a $1033\ nm$ repump pulse, pump the remaining population out of
the $|\downarrow\rangle$ or $|\uparrow\rangle$ state respectively.

State detection begins with electron shelving. A $8.5\ \mu s$ long,
$\pi$-pulse on the $|\uparrow\rangle\rightarrow D_{5/2,+3/2}$
transition is applied. To increase shelving efficiency another, $14\ \mu s$ long, $\pi$-pulse is applied on the
$|\uparrow\rangle\rightarrow D_{5/2,-1/2}$ transition. Following
shelving, an on-resonance $422\ nm$ laser light is shined on the ion
for $500\ \mu s$, during which fluorescent photons are
collected by the PMT and their time of arrival is recorded for
further analysis. Then, any population shelved to the $D_{5/2}$
level is repumped back to the ground state using a, $100\ \mu s$
long, $1033\ nm$ laser pulse. At the end of the sequence, a red
detuned $422\ nm$ light Doppler-cools the ion until the next
sequence begins. During the entire sequence the $1092\ nm$ laser is
left on to repump population from the $D_{3/2}$ metastable level.

\section{State Detection Results}

\begin{figure}[htb]
\begin{center} \subfloat[\label{fig:HistBright}]
{\includegraphics[width=0.5\textwidth]{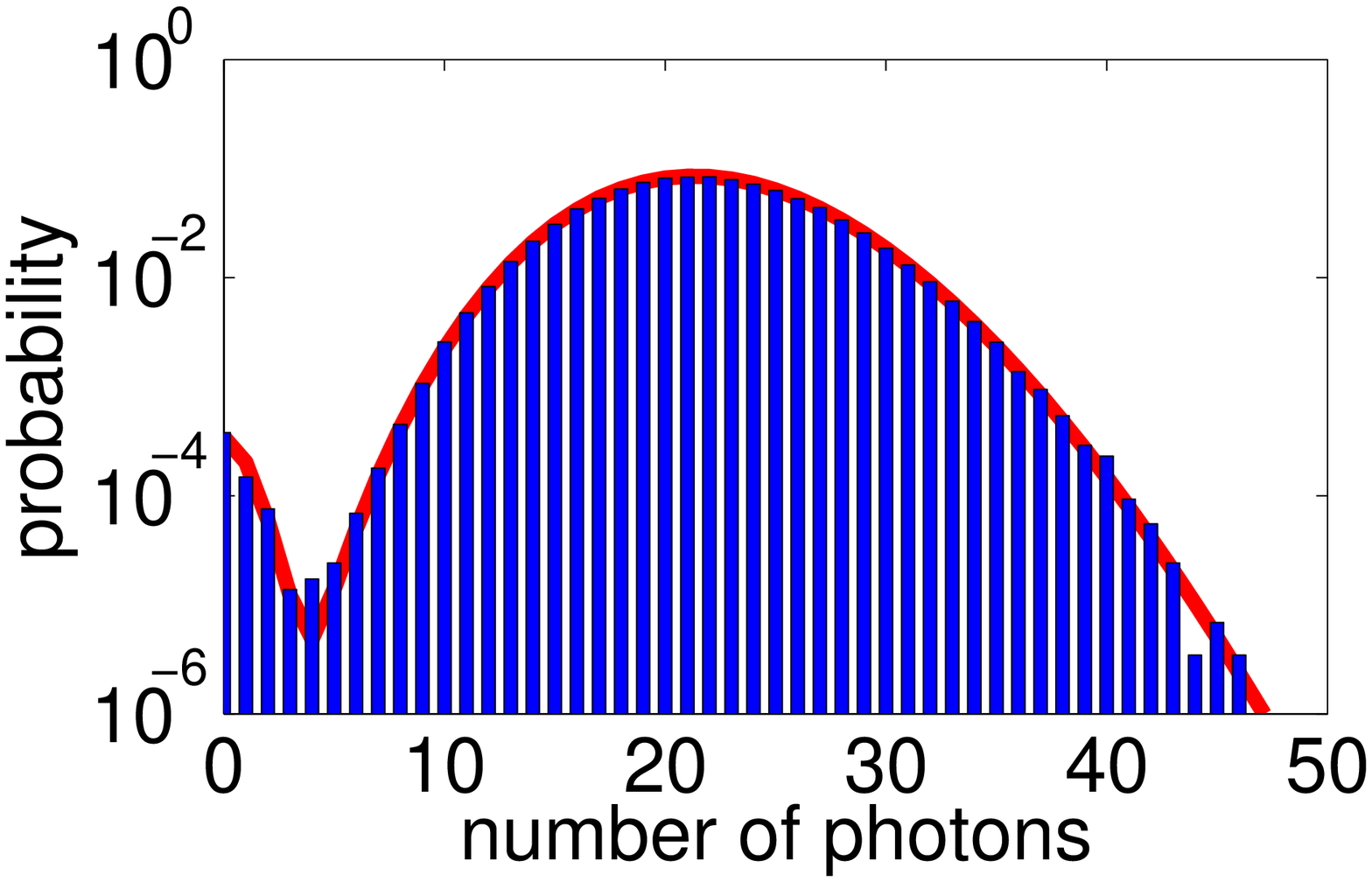}}
\subfloat[\label{fig:HistDark}]
{\includegraphics[width=0.5\textwidth]{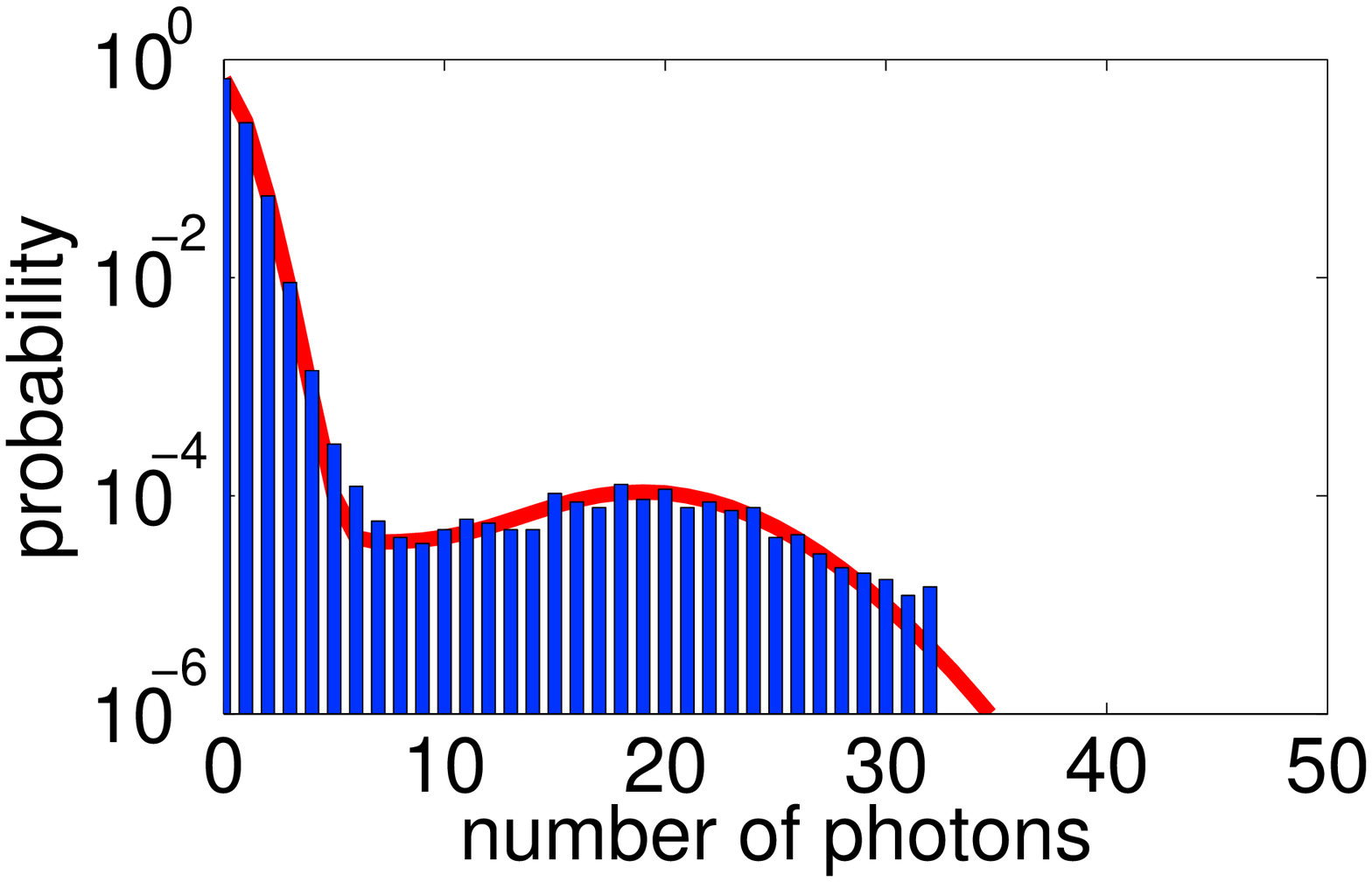}}
\caption{Experimentally obtained PDFs for a detection time of
$t_{det}=285\ \mu s$. The ion is initialized in the $|\downarrow\rangle$ state in
Fig. (a) and in the $|\uparrow\rangle$ state in Fig. (b). A maximum likelihood fit to
the expected distribution functions, given by Eqs.
\ref{eq:PDFWithErrorBrightApp} and \ref{eq:PDFWithErrorDarkApp}, are
shown (solid red line). Total initialization and shelving errors for
the $|\downarrow\rangle$ and $|\uparrow\rangle$ states obtained from the fit are
$\epsilon_{\downarrow}=6(1)\cdot10^{-4}$ and $\epsilon_{\uparrow}=10(1)\cdot10^{-4}$
respectively.}
\end{center}
\end{figure}

Normalized histograms of the number of detected photons that were
obtained in the two experiments for $t_{det}=285\ \mu s$, are
plotted in Figs. \ref{fig:HistBright} and \ref{fig:HistDark}. A
maximum likelihood fit to the expected distribution functions given
by Eqs. \ref{eq:PDFWithErrorBrightApp} and
\ref{eq:PDFWithErrorDarkApp} is denoted by the solid red line. The
known detection time $t_{det}=285\ \mu s$ and $D_{5/2}$ level
lifetime $\tau_{D_{5/2}}=390\ ms$ \cite{sinclair2005lifetime} are
used whereas $\bar{n}_{b}$, $\bar{n}_{d}$, $\epsilon_{\downarrow,tot}$ and
$\epsilon_{\uparrow,tot}$ are fit parameters. The sum of
the initialization and shelving errors for the bright and dark states
obtained from the fit are $\epsilon_{\downarrow,tot}=6(1)\cdot10^{-4}$ and
$\epsilon_{\uparrow,tot}=10(1)\cdot10^{-4}$ respectively.

To determine the minimal detection error and the optimal parameters
required to obtain it, we contour plot the measured mean error as a
function of the detection time, $t_{det}$, and the threshold value,
$n_{th}$, in Fig. \ref{fig:ExpError2D}. Here the mean error resulting from
initialization and shelving
$\frac{\epsilon_{\downarrow}+\epsilon_{\uparrow}}{2}=8(1)\cdot10^{-4}$ is
subtracted. As shown, the experimental plot reproduces the
theoretical error plot shown in Fig. \ref{fig:TheorError2D}
relatively well. In particular, the optimal parameters determined
experimentally ($t_{det}=285\mu s$, $n_{th}=6$) and the minimal
error resulting from imperfect state discrimination
$\epsilon=3(1)\cdot10^{-4}$ approach the estimated optimal
parameters ($t_{det}=280\mu s$, $n_{th}=5$), and detection error
$\epsilon=2.9\cdot10^{-4}$.
\

\section{Error Budget}

The error shown in Figures \ref{fig:TheorError2D} and
\ref{fig:ExpError2D} is fundamental and does not result from
technical imperfections. As shown by Myerson et. al.
\cite{myerson2008high}, this error can be somewhat reduced by
accounting for the photon times of arrival \footnote{By similarly
analyzing photon arrival times we were indeed able to lower this
error to $2(1)\cdot10^{-4}$}. Shelving and initialization errors
however, result from technical imperfections. In the following
discussion we try to point out the dominant sources for this errors
by calculating an order of magnitude estimates for the contribution
of different possible error sources.
A summary of this error budget is presented in Table \ref{tab:ErrorBudget}.

\subsection{Initialization Error}
The first stage of initialization consists of optical pumping to the
$|\uparrow\rangle$ state via the $P_{1/2}$ manifold. Ideally, the
optical pumping beam has to match a perfect $\sigma^{+}$
polarization. Otherwise, the matrix elements for the
$S_{1/2,+1/2}\rightarrow P_{1/2,\pm1/2}$ transitions will not null,
and some population will be pumped out of the $|\uparrow\rangle$ state.
To this end, the wave-vector, $\vec{k}$, of the beam has to be exactly
parallel to the external magnetic field and a perfect circular
polarization is required. We found the fidelity of this
initialization step in our setup to be limited to less than $0.999$
due to stress-induced birefringence in the fused silica
vacuum chamber optical ports.

To increase initialization efficiency,
we use the narrow linewidth $674\ nm$ laser to optically pump the
remaining population from the $|\downarrow\rangle$ state to the
$|\uparrow\rangle$ state via the $D_{5/2,+1/2}$ level as described
above. This initialization procedure is limited due to the finite
off-resonance excitation probability on the $S_{1/2,+1/2}\rightarrow
D_{5/2,+3/2}$ transition, resulting in a finite probability for the
electron to be in the $|\downarrow\rangle$ state at the end of the
process.
Note that there are two different contributions to
off-resonant excitation. The first is coherent with
respect to the resonant light component and is due to the pulse
finite time. The second is incoherent and is due to the servo bumps.
In steady state, i.e., after a sufficient number of pulses,
this probability is given by the ratio of the
$|\uparrow\rangle\rightarrow |\downarrow\rangle$ and
$|\downarrow\rangle\rightarrow |\uparrow\rangle$ transfer
probabilities. These are, in turn, estimated based on the
off-resonance excitation rate,
the polarization of the $1033\
nm$ repump laser, and the different decay probabilities
from the $P_{3/2}$ excited states to either the $|\uparrow\rangle$ or
$|\downarrow\rangle$ states. The incoherent off-resonance excitation rate is directly measured and is seen to be very sensitive to the specific laser and servo-loop working parameters, such as current or gain; we have observed this rate change by a factor of 2 when working with slightly different parameters. The coherent off-resonance excitation rate is estimated based on the measured Rabi frequencies and detuning.
We thus estimate the $|\uparrow\rangle$ state initialization error to
be $\sim 1\cdot10^{-4}$ due to coherent off-resonance excitation and $\sim 0.5\cdot10^{-4}$ due to incoherent off-resonance excitation.

When initializing to the $|\downarrow\rangle$ state, we initially
perform optical pumping to the $|\uparrow\rangle$ state and apply an
RF $\pi$-pulse, which transfers the electron to the $|\downarrow\rangle$
state with a fidelity higher than 0.999. This procedure is followed
by 10 similar pulses on the $S_{1/2,+1/2}\rightarrow D_{5/2,-1/2}$
transition, which result in $|\downarrow\rangle$ state initialization
with the same error as for the $|\uparrow\rangle$ state.

Note that another possible error in the initialization process could arise from leakage of $422 nm$ laser light, resulting in mixing of the two qubit states. This was, however, found not to be a problem in our system by measuring the lifetime of each of the qubit states.

\subsection{Electron Shelving Error}
The error in the shelving process is different for the two qubit states and is therefore analyzed separately.

\medskip\noindent\textit{$|\downarrow\rangle$ state shelving error.}
For the $|\downarrow\rangle$ state, the shelving error is the
probability that the ion was shelved to the $D_{5/2}$ level manifold
due to off-resonant light.
The nearest transition from $|\downarrow\rangle$ to the $D_{5/2}$ level is $5.35\
MHz$ detuned from the laser carrier, and the coupling to the
motional sidebands is suppressed by the Lamb-Dicke parameter
($\eta=0.05$ for the longitudinal motion). Here, the contribution of coherent off-resonance excitations during the two shelving pulses to the error is estimated to be $\sim 2\cdot10^{-4}$, while the error due to incoherent excitations is $\sim 1\cdot10^{-4}$.

\medskip\noindent\textit{$|\uparrow\rangle$ state shelving error.}
For the $|\uparrow\rangle$ state, the shelving error is the
probability that the ion remains in the $S_{1/2}$ manifold after
shelving was performed. A number of factors contribute to this
error.

The shelving transition Rabi frequency depends on the
ions' motion through the Debye-Waller factor. The temperature of the
ion therefore has a large effect on the shelving error. The ion is
initially Doppler cooled to a mean axial harmonic oscillator number
of $\bar{n}\simeq25$. This leads to an error of $\simeq0.01$ in a single shelving
$\pi$-pulse. It is important to note
that in the second shelving pulse the error is increased to
$\simeq0.1$ due to the change in the level occupation distribution
induced by the first pulse; following the first pulse, motional
states with a smaller Debye-Waller factor are preferentially left at
the $S_{1/2}$ level.
To reduce this error, Doppler cooling of the ion
is followed by sideband cooling on the $S_{1/2,+1/2}\rightarrow
D_{5/2,+3/2}$ transition. After five ms cooling, a mean
axial harmonic oscillator number of $\bar{n}=0.3(2)$, inferred from
the imbalance between the motional sidebands, is reached.
Yet, the mean harmonic oscillator level does not provide the full motional population
distribution, which is required to estimate the shelving error.
A thermal distribution with $\bar{n}=0.3$ would lead to a shelving error
equal to $\sim1\cdot10^{-5}$ in the first pulse and $\sim1\cdot10^{-4}$
in the second pulse. However, the distribution of high harmonic oscillator levels, following sideband cooling, is poorly described by a thermal distribution.
This is because motional states outside the Lamb-Dicke regime have a small
Debye-Waller factor on the cooling transition and therefore are not
efficiently cooled. The exact dynamics of the sideband cooling
process and the resulting energy level distribution are not
calculated here. In our estimates we use the error values assuming a thermal distribution. This is
probably an overly optimistic estimate and could be the source of the discrepancy
between our evaluated and measured errors for the $|\uparrow\rangle$ state.

Motion along the radial direction of the trap has a much weaker
effect on shelving infidelity, because of the stronger confinement along
this direction. However, while sideband cooling is performed for the axial motion,
the radial motion is only Doppler cooled and thus can not be neglected.
We estimate the infidelity due to radial motion to be $\sim3\cdot10^{-4}$ in the first pulse and
$\sim 1\cdot10^{-3}$ in the second pulse.

Intensity fluctuations of the $674\ nm$ laser are reduced to a
negligible level using an intensity noise eater. An effective
residual intensity noise resulting from beam pointing fluctuations
on the ion is estimated to give an error of $\sim3\cdot10^{-3}$ in
a single shelving pulse.

Magnetic field fluctuations are studied in a Ramsey experiment on
the $|\uparrow\rangle \rightarrow |\downarrow\rangle$ transition,
where a coherence time of $500\mu s$ is measured. The resulting errors are estimated by
numerical solution of the optical Bloch equations with a corresponding coherence decay term. Taking into  account the different magnetic field susceptibilities for the
different transitions, as well as the different durations for the
two shelving pulses, we estimate the errors to be
$\sim1.5\cdot10^{-3}$ and $\sim4.5\cdot10^{-3}$ for the two
$\pi$-pulses respectively.

A different contribution to the shelving error results from laser
frequency noise. As detailed above, this noise has few spectral contributions.
Noises that are slow compared with the experiment time
result in a typical laser frequency drift of two
$kHz$ between consecutive scans of the atomic line, giving a shelving error of $\sim3\cdot10^{-3}$ per pulse.
The contribution of the laser linewidth to the decoherence rate
is found in a Ramsey experiment on the $S_{1/2,+1/2}\rightarrow
D_{5/2,+3/2}$, yielding a coherence time of $700\ \mu s$. After
quadratically subtracting the contribution of magnetic field
fluctuations to this decoherence rate, we estimate the error owing to
the linewidth of the laser alone to be $\sim2\cdot10^{-3}$ in a
single shelving pulse.

The finite decay probability from the $D_{5/2}$ level manifold
during the shelving process also contributes to the $|\uparrow\rangle$
state shelving error. This probability has two contributions. The
first is due to the finite lifetime of the metastable
level $\tau_{D_{5/2}}=390\ ms$ \cite{sinclair2005lifetime}. After
the first shelving pulse, the electron is found at the $D_{5/2}$
level with high probability and therefore it decays with rate
$\frac{1}{\tau_{D_{5/2}}}$ during the second shelving pulse. During
the first shelving pulse this error is half as large, because, on average, only half of the population is in the
$D_{5/2}$ manifold. Using pulse durations of $8.5\ \mu s$ and $14\
\mu s$ we estimate this effect to give a contribution of
$\sim0.5\cdot10^{-4}$ to the shelving error.
The second contribution is due to off-resonant coupling of light to
motional sidebands. This coupling results in population transfer
from the $D_{5/2}$ manifold back to the $|\uparrow\rangle$ state.
The largest such contribution is due to the incoherent light in the
servo bump which has a significant spectral overlap with the
axial-motion sidebands, located $1.1\ MHz$ away from the carrier.
The estimated contribution to the error here is $\sim0.5\cdot10^{-4}$ for
the two shelving pulses combined.

In addition, leakage of $1033 nm$ laser light during detection can shorten the lifetime of the $D_{5/2}$
meta-stable state and thus reduce detection fidelity. In this experiment we find the lifetime of the $D_{5/2}$
level to be consistent with \cite{sinclair2005lifetime}, indicating that leakage of repump light has no
significant effect.

Summing up all the different contributions for the $|\uparrow\rangle$ state shelving error
after both $\pi$-pulses we get an error estimate equal to $\sim2.5\cdot10^{-4}$.

\begin{table}[htb]
  \centering
\begin{tabular}{|l|c|c|c|c|}
  \hline
  {\bf Error Source} & \multicolumn{2}{|c|}{\bm{$|\uparrow\rangle$}} & \multicolumn{2}{|c|}{\bm{$|\downarrow\rangle$}} \\
  \hline
  {\bf Initialization} & \multicolumn{2}{|c|}{ } & \multicolumn{2}{|c|}{ } \\
  \hline
  Coherent off-resonance excitation  & \multicolumn{2}{|c|}{$1$} & \multicolumn{2}{|c|}{$1$} \\
  \hline
  Incoherent off-resonance excitation  & \multicolumn{2}{|c|}{$0.5$} & \multicolumn{2}{|c|}{$0.5$ } \\
  \hline
  {\bf Initialization Total Error} & \multicolumn{2}{|c|}{\bm{$1.5$}}  & \multicolumn{2}{|c|}{\bm{$1.5$}} \\
  \hline
  {\bf Shelving} & \multicolumn{2}{|c|}{ } & \multicolumn{2}{|c|}{ } \\
  \hline
  Shelving Error Sources & $1^{st}$ pulse & $2^{nd}$ pulse & $1^{st}$ pulse & $2^{nd}$ pulse \\
  \hline
  \quad Coherent off-resonance excitation& \multicolumn{2}{|c|}{-}  & \multicolumn{2}{|c|}{$2$}  \\
  \hline
  \quad Incoherent off-resonance excitation& \multicolumn{2}{|c|}{-}  & \multicolumn{2}{|c|}{$1$}  \\
  \hline
  \quad Off-resonance sideband excitation & \multicolumn{2}{|c|}{$0.5$}  & \multicolumn{2}{|c|}{-}  \\
  \hline
  \quad $D_{5/2}$ level decay during shelving & \multicolumn{2}{|c|}{$0.5$}  & \multicolumn{2}{|c|}{-}  \\
  \hline
  \quad Axial motion in the trap &  $0.1$ & $1$ & \multicolumn{2}{|c|}{-} \\
  \hline
  \quad Radial motion in the trap &  $3$ & $10$ & \multicolumn{2}{|c|}{-} \\
  \hline
  \quad Beam pointing noise & $30$ & $30$ & \multicolumn{2}{|c|}{-} \\
  \hline
  \quad Frequency drift & $30$ & $30$ & \multicolumn{2}{|c|}{-} \\
  \hline
  \quad Fast laser linewidth & $20$ & $20$ & \multicolumn{2}{|c|}{-} \\
  \hline
  \quad Magnetic field noise & $15$ & $45$ & \multicolumn{2}{|c|}{-} \\
  \hline
  {\bf Shelving Total Error} & \multicolumn{2}{|c|}{\bm{$2.5$}}  & \multicolumn{2}{|c|}{\bm{$3$}} \\
  \hline
  {\bf Overall estimated error} & \multicolumn{2}{|c|}{\bm{$4$}}  & \multicolumn{2}{|c|}{\bm{$4.5$}} \\
  \hline
  {\bf Overall measured error} & \multicolumn{2}{|c|}{\bm{$10(1)$}}  & \multicolumn{2}{|c|}{\bm{$6(1)$}} \\
  \hline
\end{tabular}
  \caption{Shelving and initialization error budget for the two different qubit states. Errors are given in units of $10^{-4}$.
The initialization error results from off-resonant coupling to the other qubit state. The different error sources in
the shelving operation are detailed. The total shelving error for the $|\uparrow\rangle$ state is obtained by
summing all shelving error sources in each of the pulses, excluding the contributions of different off-resonance excitations and $D_{5/2}$ level decay. After the two sums are multiplied, the excluded errors are added.}\label{tab:ErrorBudget}
\end{table}

\subsection{Error Summary}\label{sec:ErrorSummary}

A summary of the estimated initialization and shelving errors and their sources is presented in Table \ref{tab:ErrorBudget} in units of $10^{-4}$. The total estimated errors for the $|\uparrow\rangle$ and
$|\downarrow\rangle$ states are $\sim4\cdot10^{-4}$ and $\sim4.5\cdot10^{-4}$ respectively, whereas the measured errors of $10(1)\cdot10^{-4}$ and $6(1)\cdot10^{-4}$ respectively are roughly twice as large.

As the error budget suggests, one significant source for initialization and shelving errors
is off-resonance excitation. This is not surprising considering that here one needs to spectrally distinguish between two states that differ by a $13\ MHz$ on top of a $445\ THz$ transition. There are two main contributions to off-resonance excitation. The first is coherent and is due to the pulse finite time, while the second is incoherent and is driven by the laser servo bumps. There are several possible avenues for reducing these errors. Increasing the magnetic field and thus the separation between the qubit states will reduce the off-resonance excitation probability. Decreasing the shelving laser intensity will reduce both coherent and incoherent off-resonance excitation errors. Slowing down the shelving transition Rabi frequency, thereby increasing the pulse length, will reduce the coherent off-resonance excitation error in proportion to the laser intensity. Second, since the pulse time is proportional to the square root of the laser intensity whereas the incoherent off-resonance excitation rate is proportional to the laser intensity, the incoherent off-resonance excitation error will reduce as the intensity square root. Note that this error reduction will come at the expense of larger error contributions due to slower laser frequency noise (drift and linewidth) and magnetic field noise. The shelving laser servo bumps can be reduced by engineering a faster servo system, by using a laser that has a narrower noise bandwidth to begin with (diode laser systems are notorious for their broad frequency noise spectrum), or by spectral filtering.

Here, incoherent off-resonance excitation errors are below $10^{-3}$. However, it is worth noting that this error source becomes much larger when carrying out operations that are off-resonance from the carrier, where the servo bump and the transition carrier have a significant overlap. We observe a large error caused by the servo bump when performing Rapid Adiabatic Passage (RAP) on the $S_{1/2}\rightarrow D_{5/2}$ transition \cite{wunderlich2007rap}, as well as when driving the motional sideband.

As mentioned above, a possible explanation for the discrepancy between the estimated and measured $|\uparrow\rangle$ state errors is a deviation of the ions' harmonic oscillator energy level distribution from a thermal distribution. Such a deviation is not surprising since in instances in which the ion motion is high, both the sideband cooling mechanism and the shelving transition are likely to fail. As an example, starting from a thermal distribution after Doppler cooling, failure to sideband cool all the instances where the ion motion is in $n>90$ will produce a larger shelving error than that we measure. Reducing this error contribution will require better cooling of the tail of the ion energy distribution, via e.g. second-sideband cooling.

\section{High-Fidelity Quantum State Tomography}

Ion-qubit state measurement is an important tool for executing many quantum
algorithms \cite{Teleportation, Teleportation_Inns,
Error_correction} and for studying different quantum processes via
state and process tomography \cite{gate_tomography,
Teleportation_tomography, Ions_Process_tomography}. In all these
cases the states that are being measured do not necessarily coincide
with the electronic eigen-basis (the two Zeeman states in the case
of a Zeeman qubit). However, all these states are related to the
electronic eigen-basis via single qubit rotations. Typically,
measurement fidelity has been characterized as the average fidelity of
state detection for the two electronic eigen-states (which are also
the measurement eigen-basis). The fidelity of single-qubit rotations
was separately benchmarked \cite{benchmark_Knill}. However, the
error introduced by rotations is not uniformly spread over all
possible input states. To this end, the experimental estimate of
state tomography fidelity, averaged over all possible single
ion-qubit states, is beneficial.

Qubit state tomography is represented by a quantum map $\epsilon$. The fidelity of state tomography of a given
pure input state $\rho_j=|\Psi_j\rangle\langle\Psi_j|$ is therefore the fidelity of this state and the
reconstructed output state $\epsilon\left(\rho_{j}\right)$,
\begin{equation}
\label{eq:Fidelity}
F=Tr\left(\epsilon\left(\rho_{j}\right)\rho_{j}\right).
\end{equation}
The output state $\epsilon\left(\rho_{j}\right)$ is reconstructed by
\begin{equation}
\label{eq:StateTomography}
\epsilon(\rho_j) = \frac{1}{2}\left(I + p_x\sigma_x + p_y\sigma_y + p_z\sigma_z\right).
\end{equation}
Here $p_x$, $p_y$ and $p_z$ are the measured projections of $\rho_j$
on the $x$, $y$ and $z$ axes correspondingly and $\sigma_j$ are the
Pauli spin operators \cite{nielsenchuang}. Note that this
definition of the fidelity is in agreement with Eq. \ref{eq:MeanError} for
the special cases of
$|\Psi_j\rangle=|\downarrow\rangle,|\uparrow\rangle$.

Here we are interested in determining the fidelity of state tomography averaged
over all possible input states. A value for the fidelity, averaged over
all possible qubit states, can be obtained by calculating an algebraic average
of the fidelities of the six pure input states $|-z\rangle =
|\downarrow\rangle$, $|+z\rangle = |\uparrow\rangle$, $|\pm x\rangle
= (|\downarrow\rangle \pm |\uparrow\rangle)/\sqrt{2}$ and $|\pm
y\rangle = (|\downarrow\rangle \pm i |\uparrow\rangle)/\sqrt{2}$
\cite{bowdrey2002fidelity}. These measurements require the ability
to initialize the qubit in different states, as well as perform
measurements in different bases. Both are achieved via qubit
rotations.

The experimental sequence performed is similar to the one previously discussed. Briefly, the ion is first ground-state cooled. Then state preparation
is performed; the qubit is first prepared in either the $|+z\rangle$
or $|-z\rangle$ state and when required is rotated to initialize the
$|\pm x\rangle,|\pm y\rangle$ states. Projection measurement
consists of mapping the measurement basis onto the $|\pm z\rangle$
basis, once again using qubit rotations, followed by shelving and
state-selective fluorescence detection. In principle a total of 18
measurements, three for each input state, are needed. However since
there is no phase information in the detection process, only nine measurements
are not redundant and were therefore performed. The measurement
uncertainties are determined by the amount of collected statistics.
For the projections on an axis that is parallel to the input state,
$2\cdot10^{5}$ repetitions were performed, yielding an error of
$2\cdot10^{-4}$. For projections on an axis that is orthogonal to
the input state $1\cdot10^{4}$ repetitions were performed which yield
an uncertainty of $1\cdot10^{-2}$. A summary of the different
measured projections and the calculated fidelities for different
input states are presented in Table \ref{tab:StateProjections}.

\begin{table}[tb]
\centering
\begin{tabular}{|c|c|c|c|c|}
  \hline
     & {\bf X projection} & {\bf Y projection} & {\bf Z projection} & {\bf Fidelity} \\
\hline
  +z & -0.011 & -0.011 & 0.9967 & 0.9984(2) \\
  -z & 0.005  & 0.005  & -0.9975 & 0.9988(2) \\
  +x & 0.9948 & 0.003 & 0.005 & 0.9974(2) \\
  -x & -0.9957 & 0.017 & 0.005 & 0.9979(2) \\
  +y & 0.017 & 0.9948 & 0.005 & 0.9974(2) \\
  -y & -0.003 & -0.9957 & 0.005 & 0.9979(2) \\
\hline
  {\bf Averaged} &  &  &  & {\bf 0.9979(2)} \\
  \hline
\end{tabular}
\caption{Summary of the state tomography results. The uncertainties in the values of the projections on axis parallel to
the initial state are $2\cdot10^{-4}$. For projections on the orthogonal axis the uncertainties are  $1\cdot10^{-2}$.
The uncertainties for the calculated fidelities are presented in the table.} \label{tab:StateProjections}
\end{table}

The $|\pm z\rangle$ measurement fidelities here are somewhat lower
than the best effort presented above. This might be due to slightly
non-optimal detection parameters \footnote{The optimal $t_{det}$ and
$n_{th}$ values depend on the photon detection rates, $R_b$ and $R_d$,
which can drift due to small changes in the laser parameters.}. The
measurement fidelities for the $|\pm x\rangle,|\pm y\rangle$ states
are generally lower than the $|\pm z\rangle$ measurement fidelities
due to larger state initialization and detection errors caused by
imperfect qubit rotations. The non-zero projections in measurement
basis orthogonal to the initialization axis are due to systematic
errors in the preparation and the measurement sequence (e.g.
slightly incorrect pulse durations). Note, however, that all the
measured state projections are within two standard deviations from
the expected value \footnote{In fact, taking into account the
imbalance in the fidelities of $\pm Z$ , the projection on an
orthogonal axis is expected to be $-2\cdot10^{-4}$}. Based on the
measurements performed, we calculate the averaged state tomography
fidelity over the entire Bloch sphere to be $\bar{F}=0.9979(2)$.

A more complete characterization of the detection process is
achieved by performing full process tomography \cite{nielsenchuang}.
We use the chi matrix, $\chi$, representation to characterize the
completely positive map representing the detection process,
\begin{equation}
\label{eq:chiRepresentation}
\epsilon\left(\rho_j\right)=\sum^4_{m,n=1}\chi_{mn}E_{m}\rho_j E_{n}^{\dagger}.
\end{equation}
Here, the fixed set $\{E_m\}^4_{m=1} = \{I, \sigma_x, i\sigma_y,
\sigma_z\}$  forms a basis for single qubit quantum maps. Ideally,
the detection process would be represented by the chi matrix
\begin{equation}
\chi_{ideal}= \left(
 \begin{array}{cccc}
   1 & 0 & 0 & 0 \\
   0 & 0 & 0 & 0 \\
   0 & 0 & 0 & 0 \\
   0 & 0 & 0 & 0 \\
 \end{array}
\right),
\end{equation}
corresponding to the Identity operation.

\begin{figure}[htb]
\begin{center}
\subfloat[\label{fig:ChiMatrix}]
{\includegraphics[width=0.45\textwidth]{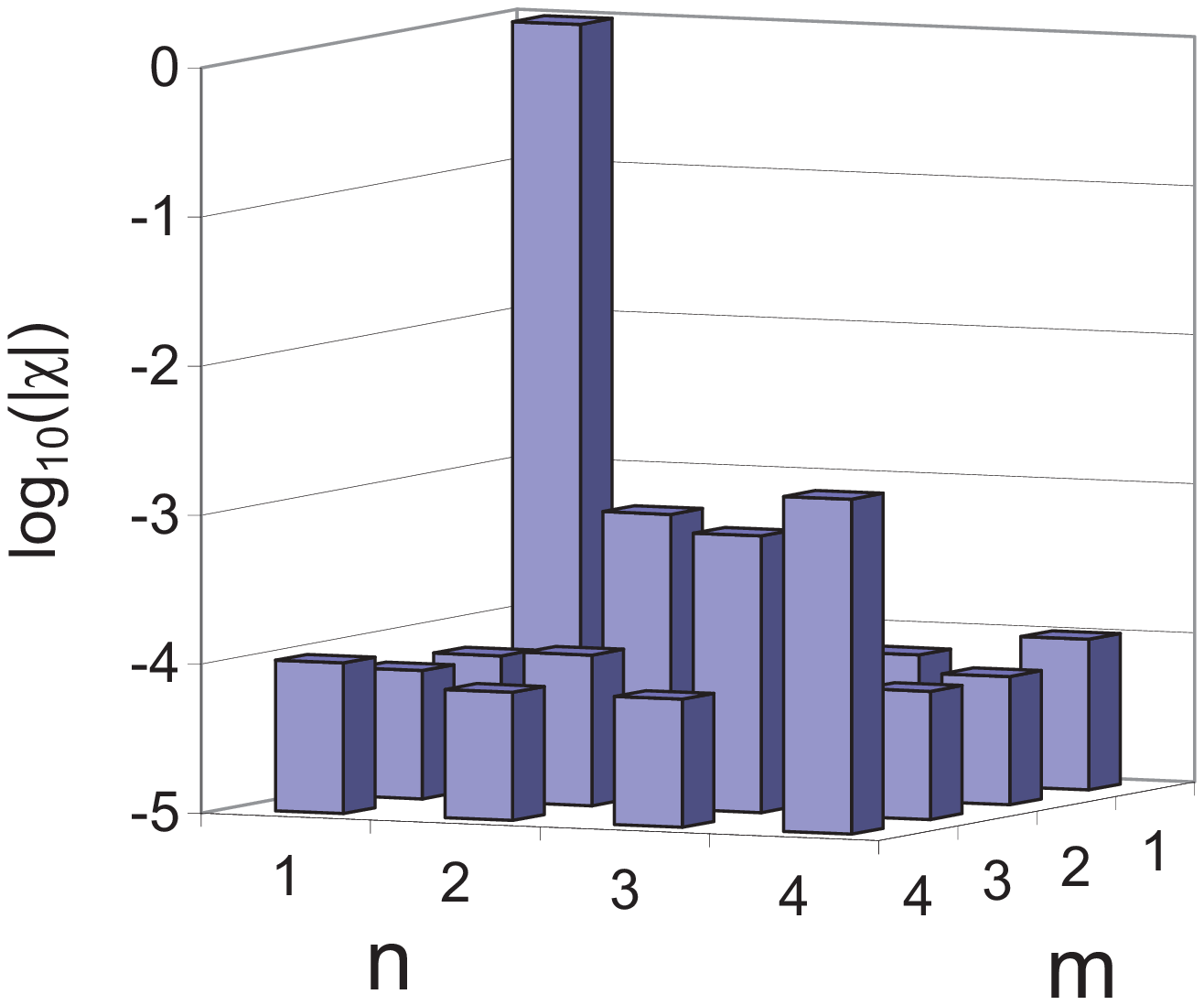}}
\subfloat[\label{fig:BlochError}]
{\includegraphics[width=0.55\textwidth]{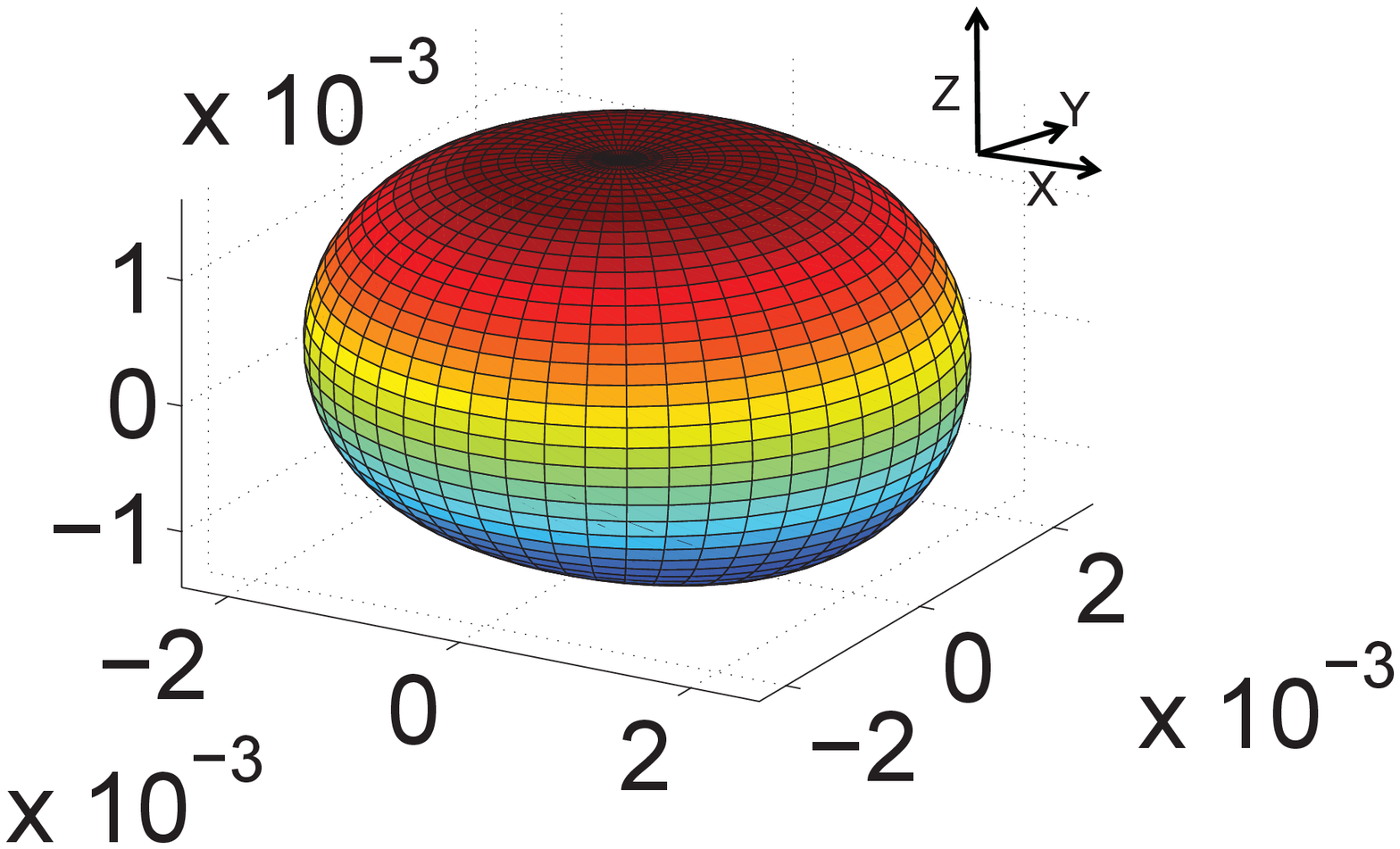}}
\caption{Process tomography results for the detection process. (a)
Absolute values of the reconstructed chi matrix entries. (b)
Detection error for every pure input state reproduced using the
reconstructed chi matrix. Here, the error in detecting the
pure state, associated with given azimuthal and polar angles on the
Bloch sphere, is represented by the radial distance from the origin
of the shown surface in that direction. The resulting spheroid is
slightly oblate due to an increased error in the equatorial
direction. This additional error is due to imperfections in the RF
pulses used to initialize qubit superpositions.}
\end{center}
\end{figure}

To determine the chi matrix of a single qubit map experimentally, it
is enough to measure the output density matrices for the following
set of four, linearly independent, input density matrices, $\{
|+z\rangle\langle +z|, |-z\rangle\langle -z|, |+x\rangle\langle +x|,
|+y\rangle\langle +y| \}$. An explicit formula can then be obtained
for the chi matrix values \cite{nielsenchuang}. The non-zero
orthogonal state projections, originating from systematic and
statistical errors, result in a non-physical (non-positive)
reconstructed chi matrix. To obtain a meaningful physical chi
matrix, we therefore null all orthogonal state projections
\footnote{A different approach would have been to perform maximum
likelihood estimate of a physical operation
\cite{Ions_Process_tomography, Maximum likelihood tomography}.
However since we believe we have identified the origin of the operation
non-positiveness with the small finite projection in the orthogonal
direction, the better approach, in our opinion, is to
null those. Two more notes: with better statistics the magnitude of
orthogonal projections can be increasingly lowered by
adjusting the rotation pulse time. Second, by artificially nulling
these projections the estimate of the error becomes larger and is
therefore pessimistic.}. The absolute values of the reconstructed
chi matrix entries are plotted in Fig. \ref{fig:ChiMatrix} on a
logarithmic scale. As expected, $\chi_{11}$, which represents the
identity operation, is three orders of magnitude larger than any
other entry. Other diagonal entries are an order of magnitude larger
than off-diagonal entries, implying that qubit depolarization is the main
error channel. The increased measurement error of superposition
states, due to rotation errors, is manifested in the slightly larger
$\chi_{44}$ (dephasing channel) as compared with $\chi_{22}$ and
$\chi_{33}$ (spin-flip channels). Off-diagonal elements are due to
the small imbalance between the measurement fidelity of the
$|+z\rangle$ and the $|-z\rangle$ states.

Using the obtained chi matrix, one can calculate the output of any input state, and hence also the fidelity. Figure \ref{fig:BlochError}
depicts the detection error over the entire Bloch sphere. The polar
angle, $\theta$, and the azimuthal angle, $\phi$, define the pure state
$|\Psi\rangle=cos\left(\frac{\theta}{2}\right)|\uparrow\rangle+e^{i\varphi}sin\left(\frac{\theta}{2}\right)|\downarrow\rangle$,
whereas the radial distance from the origin in this direction
indicates the error for this state. As expected from qubit
depolarization an almost spherical spheroid is obtained. The
resulting spheroid is also somewhat oblate due to the added error in
qubit rotations and is minutely dilated in the positive hemisphere
direction, owing to the slight imbalance between the measurement
fidelity of the $|+z\rangle$ and $|-z\rangle$ states. Using the
reconstructed chi matrix, we
find the process fidelity to be $F_{proc}=Tr(\chi_{ideal}\chi_{proc})=0.997(1)$. 

\section{Summary}

In conclusion, we demonstrate high-fidelity state detection of a
qubit encoded in the Zeeman splitting of the ground state of a
single $^{88}Sr^{+}$ ion. The limitations of the best effort readout
fidelity of $0.9989(1)$ are analyzed in detail. Our estimates of
the contribution of imperfect state preparation and shelving to the
measured error is $\sim8\cdot10^{-4}$. The remaining part,
$\sim3\cdot10^{-4}$ , results from limited state discrimination
efficiency owing to the finite lifetime of the metastable level. This
fraction of the error is intrinsic to our state detection scheme,
and can be somewhat decreased if the information on photon detection
times is used \cite{myerson2008high}.

Since one of the applications of high-fidelity state detection is performing high accuracy quantum process
tomography we measured the averaged state tomography fidelity over the entire Bloch sphere, which is $0.9979(2)$. We
also performed quantum process tomography for the detection process, and found the process fidelity to be
$F_{proc}=0.997(1)$. This fidelity can be further increased if higher fidelity qubit rotations are used.

Our measured state detection fidelity is compatible with recent estimates of the required fault-tolerance error threshold and can be used in the future to implement effective quantum error correction protocols. In addition, highly accurate quantum process tomography, important for studying basic quantum processes, can be implemented.

We gratefully acknowledge supported by the ISF Morasha program, the Crown Photonics Center and the Minerva Foundation.


\section{References}
\begin{thebibliography}{10}

\bibitem{divincenzo2001physical}
DiVincenzo~D~P 2000 The Physical Implementation of Quantum
Computation \emph{Fortschritte der Physik} \textbf{48} 771

\bibitem{knill2005quantum}
Knill~E 2005 Quantum computing with realistically noisy devices
\emph{Nature} \textbf{434} 39

\bibitem{bible}
Wineland~D~J \emph{et al} 1998 Experimental Issues in Coherent
Quantum-State Manipulation of Trapped Atomic Ions \emph{J. Res. Nat.
Inst. Stand. Tech. } {\textbf 103} 259

\bibitem{myerson2008high}
Myerson~A~H \emph{et al} 2008 High-fidelity readout of trapped-ion
qubits \emph{Phys. Rev. Lett.} \textbf{100} 200502

\bibitem{Burrell2010multi}
Burrell~A.~H. Szwer~D.~J. Webster~S.~C. and Lucas~D.~M. 2010 Scalable simultaneous multiqubit readout with $99.99\%$ single-shot fidelity
\emph{Phys. Rev. A}  \textbf{81} 040302(R)

\bibitem{qnd_detection-hume-wineland}
Hume~D~B, Rosenband~T and Wineland~D~J 2007 High-fidelity adaptive
qubit detection through repetitive quantum nondemolition
measurements \emph{Phys. Rev. Lett.} \textbf{99} 120502

\bibitem{langer2005long}
Langer~C \emph{et al} 2005 Long-lived qubit memory using atomic ions
\emph{Phys. Rev. Lett.} \textbf{95} 60502

\bibitem{cd+hyperfine_ccd}
Acton~M Brickman~K~A Haljan~P~C Lee~P~J Deslauriers~L and Monroe~C
2006 Near-Perfect Simultaneous Measurement of a Qubit Register
\emph{Quantum Inf. Comp.} \textbf{6} 465

\bibitem{langerthesis}
Langer~C~E 2006 High Fidelity Quantum Information Processing with
Trapped Ions \emph{PhD Thesis} University of Colorado

\bibitem{yb+hyperfine_detection_monroe}
Olmschenk~S, Younge~K~C Moehring~D~L Matsukevich~D~N Maunz~P and
Monroe~C 2007 Manipulation and detection of a trapped $ Yb^{+} $
hyperfine qubit \emph{Phys. Rev. A} \textbf{76} 052314

\bibitem{SchaetzPRL}
Schaetz~T \emph{et al} 2005 Enhanced Quantum State Detection
Efficiency through Quantum Information Processing \emph{Phys. Rev.
Lett.} \textbf{94} 010501

\bibitem{Oxford2004}
McDonnell~M~J \emph{et al} 2004 High-Efficiency Detection of a
Single Quantum of Angular Momentum by Suppression of Optical Pumping
\emph{Phys. Rev. Lett.} \textbf{93} 153601

\bibitem{wunderlich2007rap}
Wunderlich~C 2007 Robust state preparation of a single trapped ion
by adiabatic passage \emph{Journal of Mod. Opt.} \textbf{54} 1541

\bibitem{schmidt-kaler2009zeemanCa}
Poschinger~U~G \emph{et al} 2009 Coherent manipulation of a
$^{40}Ca^{+}$ spin qubit in a micro ion trap \emph{Journal of
Physics B} \textbf{42} 154013

\bibitem{nielsenchuang}
Nielsen~M~A and Chuang~I~L 2000 \emph{Quantum Computation and
Quantum Information} (Cambridge University Press)

\bibitem{Sengtock1994}
Sengtock~K, Sterr~U M\"{u}ller~J~H Rieger~V Bettermann~D and Ertmer~W 1994 Optical Ramsey spectroscopy on laser-trapped and thermal Mg atoms \emph{Appl. Phys. B} \textbf{59} 99

\bibitem{Yinnon's paper} Glickman~Y \emph{et al}, In preperation.

\bibitem{mythesis}
Keselman~A 2010 High Fidelity Ion Qubit State Detection \emph{MSc
Thesis} Feinberg Graduate school, Weizmann Institute of science

\bibitem{sinclair2005lifetime}
Letchumanan~V Wilson~M~A Gill~P and Sinclair~A~G 2005 Lifetime
measurement of the metastable $4d$ $^{2}D_{5/2}$ state in
$^{88}Sr^{+}$ using a single trapped ion \emph{Phys. Rev. A}
\textbf{72} 12509

\bibitem{Teleportation}
Barrett~M~D \emph{et al} 2004 Deterministic quantum teleportation of
atomic qubits \emph{Nature} \textbf{429}, 737

\bibitem{Teleportation_Inns}
Riebe~M \emph{et al} 2004 Deterministic quantum teleportation with
atoms \emph{Nature} \textbf{429}, 734

\bibitem{Error_correction}
Chiaverini~J \emph{et al} 2004 Realization of quantum error
correction \emph{Nature} \textbf{432}, 602

\bibitem{gate_tomography}
Riebe~M \emph{et al} 2006 Process tomography of ion trap quantum
gates \emph{Phys. Rev. Lett.} \textbf{97}, 220407

\bibitem{Teleportation_tomography}
Riebe~M \emph{et al} 2007 Teleportation with atoms: quantum process
tomography \emph{New J. Phys.} \textbf{9}, 211

\bibitem{Ions_Process_tomography}
Hannemann~T Wunderlich~C Plesch~M Ziman~M and Buzek~V 2009
Scrutinizing single-qubit quantum channels: Theory and experiment
with trapped ions \emph{arXiv:}0904.0923

\bibitem{benchmark_Knill}
Knill~E \emph{et al} 2008 Randomized benchmarking of quantum gates
\emph{Phys. Rev. A} \textbf{77}, 012307

\bibitem{Maximum likelihood tomography}
Ziman~M Plesch~M Bu\v{z}ek~V and \v{S}telmachovi\v{c}~P 2005 Process
reconstruction: From unphysical to physical maps via maximum
likelihood \emph{Phys. Rev. A} \textbf{72}, 022106


\bibitem{bowdrey2002fidelity}
Bowdrey~M~D Oi~D~K~L Short~A~J Banaszek~K and Jones~J~A 2002
Fidelity of single qubit maps \emph{Phys. Lett. A} \textbf{294} 258

\end{thebibliography}
\end{document}